\documentclass[aps,prl,reprint,superscriptaddress,nofootinbib,longbibliography]{revtex4-2}

\usepackage{graphicx}
\usepackage{xcolor}
\usepackage{bbold}
\usepackage{siunitx}
\usepackage{amsmath,amssymb}
\usepackage{braket}
\usepackage{soul}
\usepackage{url}
\usepackage{float}
\usepackage{microtype}
\usepackage{newtxtext}
\usepackage{newtxmath}
\usepackage{booktabs}
\usepackage{colortbl}
\usepackage{makecell}
\usepackage{subfiles}
\usepackage{xr}

\usepackage[hidelinks]{hyperref}
\hypersetup{
  colorlinks=true,
  linkcolor={blue!75!black!80!yellow},
  citecolor={blue!75!black!80!yellow},
  urlcolor={blue!75!black!80!yellow}
}

\begin{document}
\title{Topological Pumping Through a Localized Bulk in a Photonic Hofstadter System}

\author{Kyle Linn}
\thanks{These two authors contributed equally}
\affiliation{Department of Physics, The Pennsylvania State University, University Park, PA 16802, USA}

\author{Megan Goh}
\thanks{These two authors contributed equally}
\affiliation{Department of Physics, Amherst College, Amherst, MA 01002, USA}

\author{Sachin Vaidya}
\email{svaidya1@mit.edu}
\affiliation{Department of Physics, The Pennsylvania State University, University Park, PA 16802, USA}
\affiliation{Department of Physics, Massachusetts Institute of Technology, Cambridge, Massachusetts~02139, USA}

\author{Christina J{\" o}rg}
\email{cjoerg@rptu.de}
\affiliation{Physics Department and Research Center OPTIMAS, RPTU University Kaiserslautern-Landau, Kaiserslautern D-67663, Germany}

\author{Mikael C. Rechtsman}
\affiliation{Department of Physics, The Pennsylvania State University, University Park, PA 16802, USA}

\begin{abstract}
Photonic systems provide a highly tunable platform for emulating quantum Hall physics. This tunability enables probing of the interplay between strong disorder and robust topological transport that remains difficult to access in solid-state systems. Here we realize a photonic version of the Harper–Hofstadter and Aubry–André models using a one-dimensional multilayer photonic crystal (Bragg stack) with a synthetic dimension encoded in its geometry. By modulating the layer thicknesses, we observe the Hofstadter butterfly and its chiral edge states from a family of one-dimensional multilayer structures, consistent with the Thouless pump picture. Exploiting the quasiperiodicity in this model, we show that increasing quasiperiodic modulation induces a wavelength-selective localization transition: specific Chern bands become fully localized along one dimension, while chiral edge states persist and continue to wind across the gap. We confirm this behavior through numerical simulations and experiments, and eigenmode analysis reveals that edge transport in this regime proceeds via a sequence of Landau–Zener transitions between localized states. These results demonstrate a crossover from adiabatic Thouless pumping under weak quasiperiodic modulation to a Landau–Zener–mediated topological pump at strong modulation, realized in a compact and easily tunable photonic system.
\end{abstract}
\maketitle

The quantum Hall effect (QHE) provides the canonical example of topological transport, where a two-dimensional electron gas in a strong magnetic field supports chiral edge states protected by a Chern number \cite{klitzing1980new,halperin1982quantized}. These states emerge from a gapped bulk spectrum and remain robust against scattering as long as the gap remains open. Lattice generalizations, including Chern insulators and Thouless pumps, demonstrate that quantized transport can arise in periodic systems \cite{thouless1983quantization}. Analogous phases have been realized in topological photonics \cite{lu_topological_2014,ozawa_topological_2019,https://doi.org/10.1002/lpor.202100300,10.1063/5.0058478}, where effective time-reversal symmetry breaking enables robust light transport \cite{raghu2008analogs, wang2009observation, rechtsman2013photonic, hafezi2013imaging, jin2023observation}. This robustness underpins potential photonic applications including topological information transport \cite{hafezi_robust_2011}, optical isolation \cite{zhou2017optical, el2013chip, el2015optical}, and topological lasing \cite{doi:10.1126/science.aar4003,doi:10.1126/science.aar4005, st2017lasing}.

Because chiral edge states are valued for their robustness, the effects of weak perturbations such as fabrication imperfections have been widely studied \cite{klitzing1980new,chang2013experimental,wang2009observation,rechtsman2013photonic, hafezi2013imaging, Mittal2014TopologicallyRobust,stutzer2018photonic, yao2024observation}. By contrast, the interplay between strong perturbations and topology remains less explored. For example, strong random disorder can induce Anderson localization, collapsing extended bulk states and suppressing transport \cite{anderson1958absence}, although such disorder can also generate nontrivial phases such as topological Anderson insulators \cite{li2009topological,groth2009theory,stutzer2018photonic}. Alternatively, quasiperiodic perturbations offer a controlled route to localization physics. In the Aubry–Andr\'e (AA) model, quasiperiodic modulation produces a self-duality that determines whether bulk eigenstates are extended or localized \cite{aubry1980analyticity}. Extensions of the model can offer more control with additional phenomena such as mobility edges and reentrant delocalization transitions, observed in tight-binding chains \cite{PhysRevB.83.075105, PhysRevLett.104.070601, li2017mobility, PhysRevLett.126.106803, roy2022critical, padhan2022emergence}, cold atoms \cite{roati2008anderson, PhysRevLett.120.160404, PhysRevLett.122.170403}, and photonics \cite{kraus2012topological, PhysRevLett.103.013901, vaidya2023reentrant}. However, experimental solid-state platforms rarely allow controlled disorder or quasiperiodic modulation combined with quantum Hall physics.

Here we leverage the tunability of photonic systems to study the interplay between quasiperiodic modulation and quantum Hall transport. We use a family of one-dimensional photonic multilayer thin-films (Bragg stacks) with a synthetic dimension encoded in their geometry. By modulating the dielectric layer thicknesses, these structures map onto a higher-dimensional parameter space that reproduces the two-dimensional Harper-Hofstadter model, which includes its fractal butterfly spectrum and chiral edge states.  The AA model gives us access to two different regimes: one where the bulk states are extended throughout the whole system (akin to the extended Bloch states of Chern insulators), and one where they are localized (akin to the localized Landau level eigenstates associated with the quantum Hall effect in homogeneous space).  This raises the question of how photonic topological pumping proceeds differently in these two different regimes.  We find that while pumping is mediated by adiabatic transport over bulk states in the former case (as in Ref. \cite{kraus2012topological}), in the latter case, transport instead proceeds through Landau–Zener–type transitions between localized states.

\begin{figure}[]
    \centering
    \includegraphics[width=\linewidth]{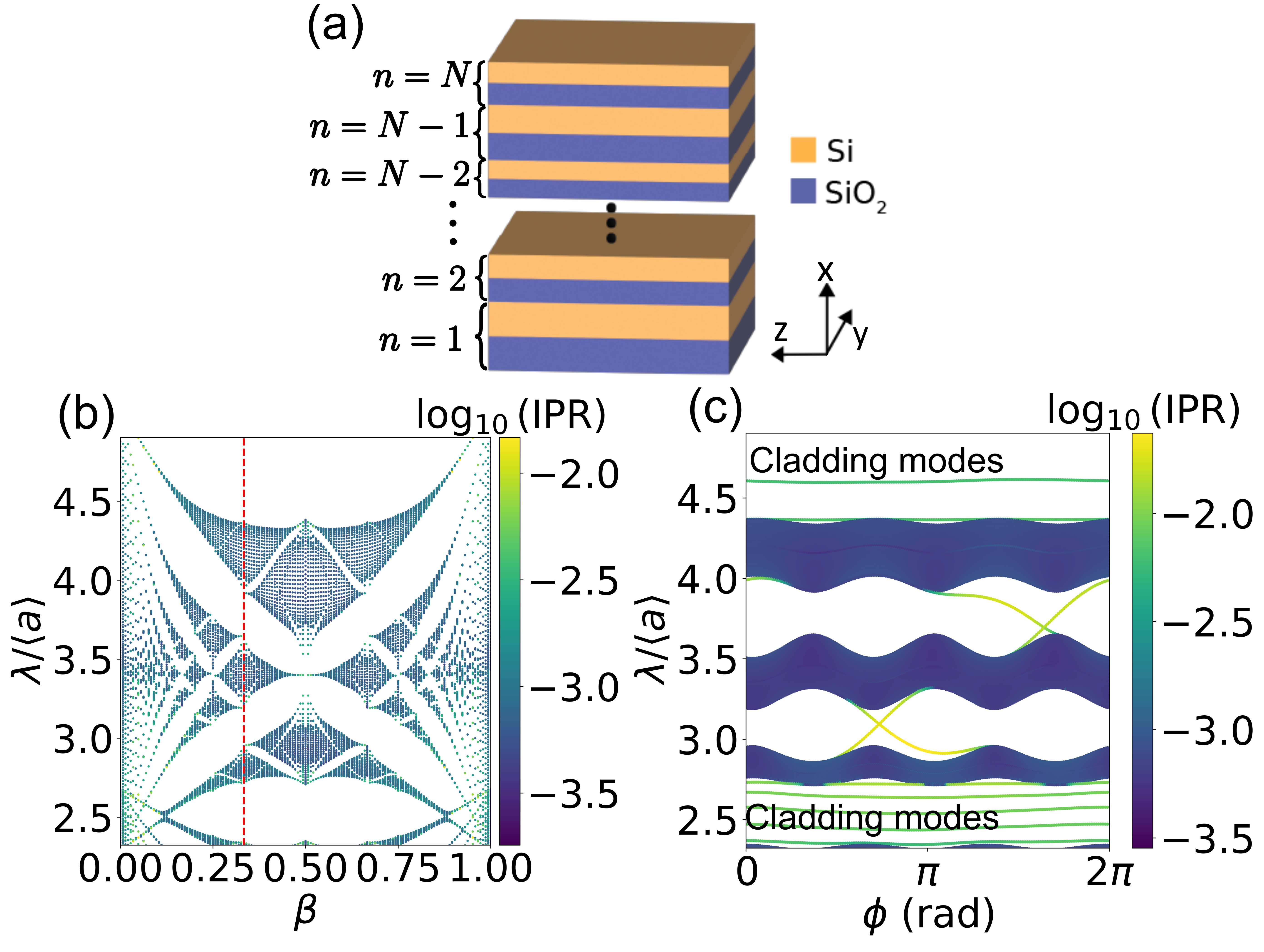}
    \caption{ \small{(a) A schematic of a 1D multilayer PhC consisting of alternating dielectric materials of Si and SiO\(_2\). An effective unit cell \(n\) is defined as a pair of neighboring Si and SiO\(_2\) layers. Within a unit cell, the layers share the same thickness \(t_n\). (b) The Hofstadter butterfly plotted by normalized wavelength and colored by \(\log_{10}\)(IPR) realized in a family of 1D PhCs of \(N=144\) with \(\beta\) as a tuning parameter. Fractal butterflies can be seen for \(\beta\) away from the middle (\(\beta = 0.50\)). A red dashed line indicates the slice of \(\beta\) fixed to realize: (c) The chiral edge states colored by \(\log_{10}\)(IPR) calculated from a family of 1D PhCs of \(N=144\) with \(\phi\) as a tuning parameter. This chosen slice of \(\beta\) slices through three extended bulk bands, and localized edge states traverse the gaps and wind once within a period of \(\phi\). Near-flat cladding modes pad the top and bottom of the spectra.
      \label{fig:fig 1}}}
\end{figure}

Our platform is a one-dimensional multilayer photonic crystal (PhC) composed of alternating dielectric layers, e.g. Si and SiO\(_2\), stacked along the \(x\) direction, as shown in Fig.~\ref{fig:fig 1}(a). We define an effective unit cell as a pair of adjacent Si and SiO\(_2\) layers. Light propagation in this structure is governed by the 1D Maxwell eigenvalue equation \cite{photoniccrystalsbook}:

\begin{equation}
    \label{maxwell}
    -{\partial_{x}}\left(\frac{1}{\varepsilon(x)}\partial_{x}\right){\mathcal{H}(x)} = \left(\frac{\omega}{c}\right)^2{\mathcal{H}(x)}
\end{equation}
Here, \(\varepsilon(x)\) denotes the dielectric function, \(\omega\) the eigenfrequency of light, \(c\) the speed of light, and \(\mathcal{H}(x)\) the scalar magnetic field of a transverse component. Because we consider only normally incident light, the two transverse polarizations (transverse electric and transverse magnetic) are degenerate and are therefore governed by the same equation. Consequently, we do not distinguish between them. When \(\varepsilon(x)\) is periodic, Bloch eigenmodes are produced through wave interference from the dielectric contrast, and their eigenfrequencies organize into photonic bands. Adjusting the unit-cell geometry (e.g. layer thicknesses) changes the optical path and dielectric profile within each period, shifting band frequencies in a way analogous to tuning the on-site energies in a 1D Schrödinger model. In this sense, the geometric modulation provides a controlled way to shape the effective lattice potential experienced by an optical field.

Inspired by the on-site modulation of energies in the Harper-Hofstadter model \cite{hofstadter1976energy}, we tune the thicknesses of each layer within a unit cell with the following equation:

\begin{equation} \label{thickness eq}
    t_n = t_0\left[1 + A\cos(2\pi \beta n + \phi)\right]
\end{equation}
Here, \(n \in \{1,2,\ldots , N\}\) labels a unit cell that contains two layers, Si and SiO\(_2\), of equal thickness \(t_n\). \(t_0\) is an initial thickness which sets a length scale, so that the average lattice constant (distance between pairs of Si and SiO\(_2\)) is \(\braket{a} = 2t_0\). Additionally, \(A\) is the amplitude of modulation bounded by \(|A| \leq 1\), \(\beta\) is the (spatial) frequency of modulation akin to the magnetic flux in the electronic case, and \(\phi\) is an added degree of freedom which manifests as an offset in thicknesses in the layers. Comparing Eq.~\eqref{thickness eq} to a standard Harper equation (reviewed in the supplementary), \(\phi\) maps onto the transverse momentum in the electronic case. This extra parameter \(\phi\) thus acts as a synthetic dimension, for which edge states are dispersive in this parameter space but live localized in real-space. This use of synthetic dimensions is common \cite{ozawa2019synthetic, DUTRA2023114338, lustig2019edge} as a powerful framework to access physics beyond a strictly 1D physical system \cite{maczewsky2020synthesizing, PhysRevLett.132.266601}.

\begin{figure*}[t]
    \centering
    \includegraphics[width=\textwidth]{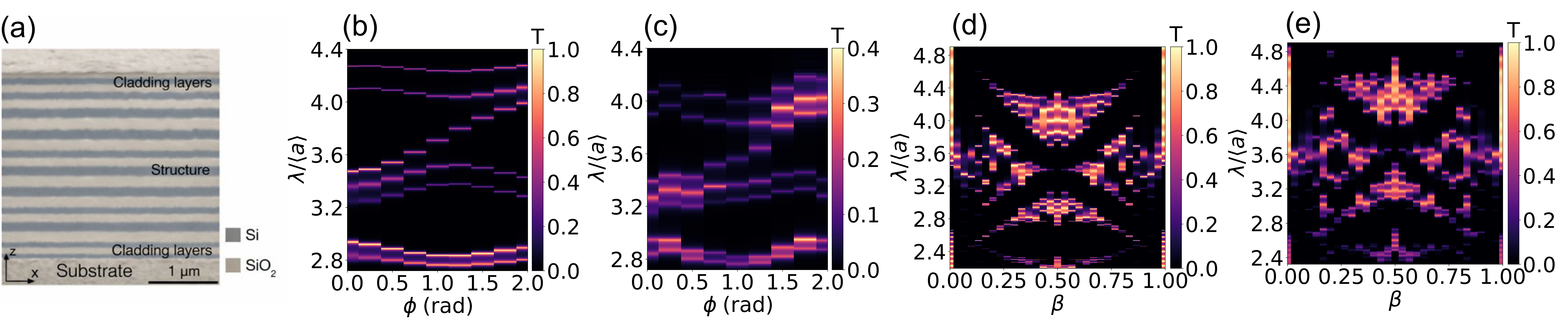}
    \caption{ \small{ (a) SEM image of an \(N=8\) 1D PhC including cladding layers used for characterizing chiral edge states. The multilayer stack is composed of Si (dark grey) and SiO\(_2\) (light grey) deposited onto a borosilicate glass cover slide as the substrate. Cladding layers on each side of the PhC extend the edge states' lifetime. (b) Transfer-matrix simulation of chiral edge states within the bandgap at \(\beta=1/3\), \(A = 0.2\), for an \(N=8\) system. An edge branch crosses the top bandgap. (c) Experimentally measured transmission spectrum showing the same edge state. (d) Transfer-matrix simulation of Hofstadter butterfly in an \(N=13\) system. (e) Experimentally measured transmission showing the butterfly.
      \label{fig:fig 2}}}
\end{figure*}

We assemble a family of 1D PhCs, where each individual PhC has a different set of \(t_n\), such that varying \(\beta\) and \(\phi\) across the family reconstructs different topological features. For example, to reproduce the Hofstadter butterfly, which is originally the eigenvalue spectrum as a function of magnetic flux, we tune \(\beta\) across different PhCs while keeping \(\phi\) fixed. We then compute the spectrum of each individual PhC from Eq.~\eqref{maxwell} and stitch together the spectra to form a photonic Hofstadter butterfly. A similar procedure reproduces the chiral edge states. In the electronic problem, one first selects a gap of the Hofstadter butterfly by fixing \(\beta\), then obtains the corresponding edge-state band structure by imposing open boundary conditions and calculating eigenvalues as a function of transverse momentum. In our case, this is realized by tuning \(\phi\) across a family and fixing \(\beta\). Thus, the resolution in each of the resulting photonic spectra is set by the number of PhCs, where each slice is a separate multilayer stack.

We utilize a plane-wave expansion method through MIT Photonic Bands (MPB) software to calculate the photonic eigenmodes, i.e. the solutions to Eq.~\eqref{maxwell} \cite{johnson2001block}. A resulting photonic Hofstadter butterfly is shown in Fig.~\ref{fig:fig 1}(b) through a plot of eigenstates at normalized wavelength vs. \(\beta\), with each eigenstate colored by the logarithm (base 10) of their Inverse Participation Ratio (IPR). The IPR diagnoses the degree of a state's localization, with higher values (yellow) indicating localized states and lower values (purple) indicating extended states. The fractal pattern similar to tight-binding systems can be seen in this case. A red dashed line on the plot indicates \(\beta = 1/3\), for which we keep fixed and sweep Eq.~\eqref{thickness eq} for \(\phi\) to see the chiral edge states depicted in Fig.~\ref{fig:fig 1}(c). 

To see these edge states, the system must have open boundary conditions with additional Si and SiO\(_2\) layers acting as cladding. Since light is not confined to a material as electrons are by a work function, this cladding extends the lifetime of states living on the edge, allowing them to be imaged in both simulation and experiment. The additional modes in the spectrum (of Fig.~\ref{fig:fig 1}(c)) that exist above and below the bulk bands (\(\lambda/\braket{a} \gtrsim 4.3\) and \(\lambda/\braket{a} \lesssim 2.7\)) are modes from this cladding. The bulk bands in the spectrum are highly delocalized (extended) while there exist chiral edge states traversing the gap that are dispersive in the synthetic dimension \(\phi\). The IPR indicates that these edge states are localized, implying that they live on the physical edges of each 1D PhC. 

The localization of a photonic eigenstate implies there is no channel of transmission of light through the system, provided that the system size is larger than the localization length. Meanwhile, extended states imply that transmission of light persists regardless of the system size. We can thus confirm these spectra in experiment through transmission measurements of fabricated 1D PhCs and begin discussing how quasiperiodic modulation can impact transport.

\begin{figure*}[t]
    \centering
    \includegraphics[width=\textwidth]{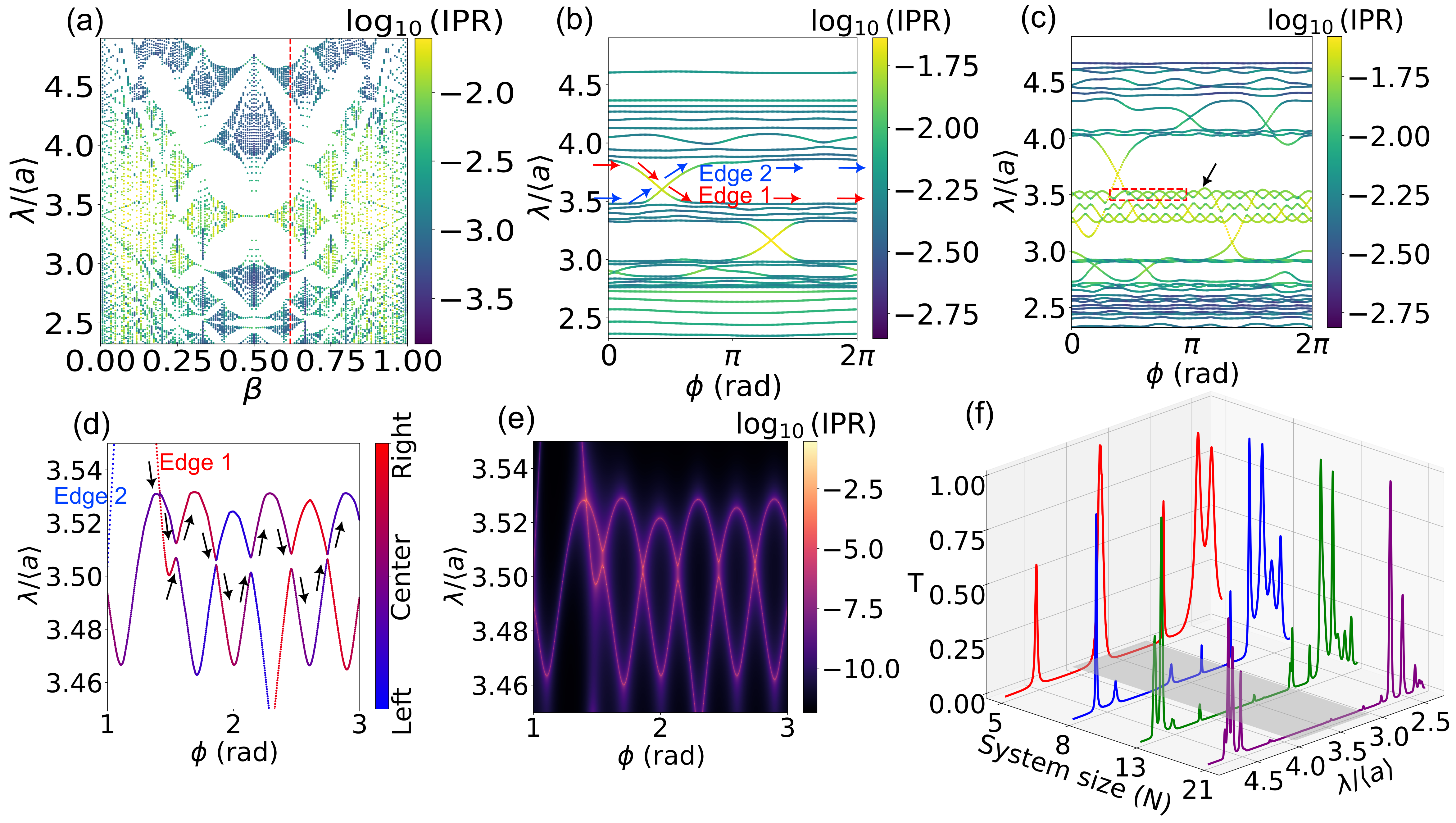}
    \caption{ \small{ (a) MPB simulation of the photonic Hofstadter butterfly at \(N=144\) and high modulation \(A=0.5\). The simulation shows pockets of localization that are wavelength- and \(\beta\)-dependent. The red dashed line indicates the inverse golden ratio, \(\beta \approx 2/(\sqrt{5}+1)\), which is the slice chosen to reproduce the chiral edge states in (b) and (c). (b) MPB simulation of the chiral edge states at \(N=21\), \(\beta = 34/21\), and at low modulation \(A = 0.2\). Nearly all modes are extended except the chiral edge states, which are localized at the edges of the 1D PhQC. An edge branch can be smoothly tracked as it enters the bulk and wraps around to the other branch within a period of \(\phi\), indicated by the red arrows from Edge 1 that wrap onto the blue arrows at Edge 2.  (c) The same as (b) except at higher modulation \(A=0.5\). The bulk bands located near \(\lambda/\braket{a} \sim 3.5\), indicated by the black arrow, are localized, while the other bands remain extended. Chiral edge states continue to cross the gap in this case. (d) Center-of-mass calculations in real space of the energy density for eigenstates highlighted in the red dashed box in (b). A possible eigenstate's evolution starts at the right edge (Edge 1) and traverses multiple avoided crossings indicated by the color. (e) Transfer-matrix simulations of the logarithmic transmission of the same region of (d), showing avoided crossings. (f) Experimental results for \(\phi = 1.25\), showing extended bulk bands (at the edges of the spectrum) and the localizing bulk bands and chiral edge states (highlighted by the grey shadow). Faster attenuation of the edge states and bulk bands with increasing system size, relative to the extended bulk bands, is a signature of localization.
      \label{fig:fig 3}}}
\end{figure*}

To demonstrate the existence of a Hofstadter butterfly and chiral edge states in our 1D PhC with a synthetic dimension, we fabricate several Si-SiO\(_2\) multilayer PhCs patterned by Eq.~\eqref{thickness eq} using plasma-enhanced chemical vapor deposition (PECVD) and measure the transmission using a UV-Vis-NIR spectrophotometer. We compare these results to transfer-matrix simulations, which calculate the transmission coefficients layer-by-layer through Fresnel equations \cite{yeh1990optical}. A scanning electron microscope (SEM) image of an example PhC used to replicate chiral edge states is shown in Fig.~\ref{fig:fig 2}(a), which contains the multilayer structure with cladding layers deposited onto a substrate.

In Fig.~\ref{fig:fig 2}(b), we simulate the transmission through the transfer-matrix method of several PhCs with different \(\phi\), each having a fixed \(\beta = 1/3\). The experimentally measured transmission is shown in Fig.~\ref{fig:fig 2}(c), which is in strong agreement with the simulation. One can see an edge state traversing the top gap with positive dispersion (of synthetic momentum \(\phi\)). The variations in the experimental spectrum can be explained by fluctuations in the deposition rate between different samples. Additionally, by fixing \(\phi\) and fabricating several PhCs of different \(\beta\), we reconstruct the Hofstadter butterfly shown in Fig.~\ref{fig:fig 2}(d) and (e), which are the simulation and experiment, respectively. At this resolution (i.e. number of samples), only the lowest order bandgaps, corresponding to the largest gaps near \(\beta = 0.50\), of the butterfly can be resolved, which is visible in both simulation and experiment. This confirms that the 1D PhC platform with synthetic dimensions can capture the 2D tight-binding Hofstadter model and QHE phenomena well.

With this platform's ease of tunability, we can now investigate the effects of strong perturbations and a localized bulk on this topological system. We note that the previous results were done at low amplitude of modulation (\(A = 0.2\)), which corresponds to the usual Hofstadter model. Increasing the amplitude of modulation \(A\) and calculating through MPB the resulting butterfly results in the plot shown in Fig.~\ref{fig:fig 3}(a). Here, the fractal pattern persists, though now pockets of eigenstates in the spectrum become localized. This localization occurs only for certain values of \(\beta\) and wavelength. In fact, only irrational \(\beta\), as known in standard AA models, will possess such a transition. Since irrational values are dense in the reals, the regions of localization within the spectra are dense as well. We highlight one particular value of $\beta$, marked by the dashed red line, to focus on: the inverse golden ratio \(\beta = 2/(\sqrt{5} +1)\) (approx. 0.618). This number can be approximated by successive ratios of the Fibonacci sequence. Because of the slow convergence of these rational approximants, it is known as the most irrational of the irrationals. This allows us to see the localization effects most prominently in the numerics and prevents aperiodic artifacts as long as the system size \(N\) matches the denominator in the rational approximant for \(\beta\).

Figure \ref{fig:fig 3}(b) and (c) depict chiral edge states for \(N=21\) and rational approximant \(\beta = 34/21 \) at low modulation (\(A = 0.2\)) and high modulation (\(A = 0.5\)) from the red dashed line pictured in Fig.~\ref{fig:fig 3}(a). These are now photonic quasicrystals (PhQCs), due to their well-approximated irrational frequency of modulation. As a result, the parameter \(A\) now represents the strength of quasiperiodic modulation.  Between the two modulation cases, we see a wavelength-selective localization transition that targets specific bulk bands. These selective transitions are a characteristic of mobility edges, induced by the effective long-range couplings of a photonic platform that break the self-duality in standard AA models. In the low modulation case (Fig.~\ref{fig:fig 3}(b)), all bulk bands are extended with chiral edge states traversing across the bandgaps. These edge states can be clearly seen entering an extended bulk and wrapping around to the other branch, consistent with the photonic Thouless pumping picture of Ref. \cite{kraus2012topological}. Specifically, an eigenstate originating on Edge 1 evolves continuously in momentum space (red arrows), entering the extended bulk bands near \(\lambda/\braket{a} \sim 3.5\), before wrapping from \(\phi : 2\pi \to 0\) and reconnecting to the degenerate branch on Edge 2, as indicated by the blue arrows. On the other hand, in the high modulation case (Fig.~\ref{fig:fig 3}(c)), the bulk bands near \(\lambda/\braket{a} \sim 3.5\), indicated by the black arrow, are now strongly localized. The smooth edge-to-edge connectivity seen earlier breaks down, and the synthetic momentum-space trajectory is now more complex. We note that the butterfly and chiral edge states also exhibit a reentrant delocalization transition at even higher \(A\), consistent with the findings in \cite{vaidya2023reentrant}, though we choose to focus on this first localization transition only.

In previously studied photonic Thouless pumps, bulk states are extended and pumping proceeds by following a single bulk state at the band edge that connects to the edge state \cite{kraus2012topological}.  This is strongly akin to a Thouless pump that is dimensionally reduced from a two-dimensional Chern insulator.  Since Fig.~\ref{fig:fig 3}(c) indicates a fully localized bulk, which can also be confirmed with transfer-matrix simulations at large system size (see supplementary), the Thouless pump picture based on smooth spectral flow breaks down.  The localized states observed here are much more reminiscent of Landau level eigenstates (in the Landau gauge) in the quantum Hall effect in a homogeneous medium under a magnetic field. As a result, another mechanism is needed to explain how transport of these chiral edge states can still persist, i.e., how a particle can exchange from one edge to another through a localized bulk.

To answer this question, we calculate the real-space center-of-mass of the energy density, \(x_\text{COM} = \sum_i x_i \left|\mathcal{H}(x_i)\right|^2 / \sum_i \left|\mathcal{H}(x_i)\right|^2\) where \(x_i\) is a discretized grid point in MPB, of the magnetic field eigenstates calculated for Fig.~\ref{fig:fig 3}(c). This center-of-mass calculation is shown in Fig.~\ref{fig:fig 3}(d), which is a zoomed-in view of the red dashed box shown in Fig.~\ref{fig:fig 3}(c). The states are colored by their position in the 1D PhQC, with red indicating the right edge (arbitrarily chosen), blue indicating the left edge, and purple in the middle of the system. With this method, we can roughly track how an eigenstate propagates through the bulk if we treat \(\phi\) as our pump parameter. In Fig.~\ref{fig:fig 3}(d), one possible trajectory begins at the right edge (labeled Edge 1) and moves lower in wavelength as a function of \(\phi\) until it encounters a mini-gap. These gaps are avoided crossings, caused by spatial overlaps of adjacent localized eigenstates. Intuitively, such overlaps are more naturally facilitated in PhC platforms, where eigenstates can be more spatially extended than those of standard tight-binding models with only short-range couplings.  The color of the eigenstates indicates that a particle on this trajectory (following the black arrows) must transfer across the gap, as the branches opposite the gap have nearly degenerate real-space position and synthetic momentum. 

Such exchanges across these avoided crossings are explained by Landau-Zener tunneling, which describes how an initial eigenstate in a two-level system can transition across an avoided crossing non-adiabatically \cite{landau1932theorie, zener1932non}. We emphasize that each \(\phi\) labels a distinct and static photonic crystal, meaning there is no underlying temporal dynamics or adiabatic evolution in the usual sense. Instead, the Landau-Zener framework is used here in a structural sense, as these avoided crossings mediate hybridization between nearly degenerate localized modes in an analogous manner. This pattern of avoided crossings repeats until after a full cycle of \(\phi\) (beyond the region shown) a particle returns to the blue edge within the top gap, labeled Edge 2. A transfer-matrix simulation of the transmission of this zoomed-in region, depicted in Fig.~\ref{fig:fig 3}(e), confirms the existence of these avoided crossings. 

Because the localization lengths of the bulk bands near \(\lambda/\braket{a} \sim 3.5\) and chiral edge states are small, the transmission obtained in transfer-matrix simulations is low if not absent for certain system sizes. Thus, it is impossible to directly image these avoided crossings in transmission experiments. Instead, we look to the localization transition itself as an experimental observable. This selective localization transition is confirmed experimentally by fabricating chiral edge state samples at high modulation, \(A = 0.5\), golden ratio approximant \(\beta\), and a chosen slice of \(\phi = 1.25\), which captures both edge states and bulk bands (both extended and localized). We plot the transmission as a function of system size \(N\), for four different PhQCs (\(N=5, 8, 13, 21\) which have \(\beta = 8/5, 13/8, 21/13, 34/21\) respectively) shown in Fig.~\ref{fig:fig 3}(f). In this plot, the transmission spectra are normalized to a maximum within each sample size. The extended bulk bands shown at the edge of the wavelength range (\(\lambda/\braket{a} \gtrsim 4.3\) and \(\lambda/\braket{a} \lesssim 3.0\)) show relatively high transmission which can attenuate from material loss as the system size increases. However, the edge states and localized bulk bands in between, which are highlighted by the gray shadow, drop off more sharply. Because of finite-size effects, e.g. for \(N=5\), one sees fewer peaks in transmission within the gap because of missing bulk bands, but the localization transition can still be clearly seen; the transmission of localized bands predicted in MPB drops off sharply with system size.

In conclusion, we demonstrate a simple and tunable 1D multilayer PhC that realizes higher-dimensional QHE physics through synthetic parameters. Our platform faithfully reproduces the Harper-Hofstadter model, including the fractal butterfly spectrum and its chiral edge states, and enables controlled exploration of strong quasiperiodic modulation. We observe a localization transition in which bulk Chern bands become fully localized while edge transport persists, accompanied by a crossover from adiabatic pumping via bulk states at weak quasiperiodicity to Landau-Zener-mediated transport at strong quasiperiodicity. MPB simulations, eigenmode evolution, and IPR analysis confirm bulk localization, while system-size dependent transmission provides an experimental proxy once localization lengths become too short for direct spectral resolution.

While similar phenomena have been reported in bulk-localized transport states induced by magnetic aperiodicity \cite{johnstone2022bulk} and mobility-edge-assisted pumping in quasiperiodic lattices \cite{xing2021adiabatic,xing2022quantum}, our observed effect is distinct: the bulk bands are fully localized in real space, yet transport persists through repeated Landau–Zener handoffs across narrow avoided crossings between localized states. The synthetic dimension supplies connectivity in parameter space, allowing delocalization along the synthetic direction even when real-space states are localized.

\textit{Acknowledgments}---We thank Zeyu Zhang for SEM images of the PhCs and Bill Mahoney for technical help with the PECVD process.
C.J. acknowledges funding from the Alexander von Humboldt Foundation within the Feodor-Lynen Fellowship program. 
We acknowledge the U.S. Office of Naval Research under Grant No. N00014-20-1-2325, the Air Force Office of Scientific Research under Grant No. FA9550-22-1-0339 as well as the U.S. Army Research Office under Grant Nos. W911NF-22-2-0103 and W911NF-24-1-0224. K. L. and M. G. acknowledge support from the NSF-REU Program under grant number DMR-1851987.

\nocite{peierls_zur_1933, thouless1982quantized, hatsugai1993}

\bibliography{references}

@article{raghu2008analogs,
  title={Analogs of quantum-Hall-effect edge states in photonic crystals},
  author={Raghu, Srinivas and Haldane, Frederick Duncan Michael},
  journal={Physical Review A},
  volume={78},
  number={3},
  pages={033834},
  year={2008},
  publisher={APS}
}

@article{wang2009observation,
  title={Observation of unidirectional backscattering-immune topological electromagnetic states},
  author={Wang, Zheng and Chong, Yidong and Joannopoulos, John D and Solja{\v{c}}i{\'c}, Marin},
  journal={Nature},
  volume={461},
  number={7265},
  pages={772--775},
  year={2009},
  publisher={Nature Publishing Group UK London}
}

@article{rechtsman2013photonic,
  title={Photonic Floquet topological insulators},
  author={Rechtsman, Mikael C and Zeuner, Julia M and Plotnik, Yonatan and Lumer, Yaakov and Podolsky, Daniel and Dreisow, Felix and Nolte, Stefan and Segev, Mordechai and Szameit, Alexander},
  journal={Nature},
  volume={496},
  number={7444},
  pages={196--200},
  year={2013},
  publisher={Nature Publishing Group UK London}
}

@article{hafezi2013imaging,
  title={Imaging topological edge states in silicon photonics},
  author={Hafezi, Mohammad and Mittal, Sunil and Fan, J and Migdall, A and Taylor, JM},
  journal={Nature Photonics},
  volume={7},
  number={12},
  pages={1001--1005},
  year={2013},
  publisher={Nature Publishing Group UK London}
}

@article{jin2023observation,
  title={Observation of Floquet Chern insulators of light},
  author={Jin, Jicheng and He, Li and Lu, Jian and Chang, Lin and Shang, Chen and Bowers, John E and Mele, Eugene J and Zhen, Bo},
  journal={arXiv preprint arXiv:2304.09385},
  year={2023}
}

@article{hofstadter1976energy,
  title={Energy levels and wave functions of Bloch electrons in rational and irrational magnetic fields},
  author={Hofstadter, Douglas R},
  journal={Physical review B},
  volume={14},
  number={6},
  pages={2239},
  year={1976},
  publisher={APS}
}

@article{thouless1982quantized,
  title={Quantized Hall conductance in a two-dimensional periodic potential},
  author={Thouless, David J and Kohmoto, Mahito and Nightingale, M Peter and den Nijs, Marcel},
  journal={Physical review letters},
  volume={49},
  number={6},
  pages={405},
  year={1982},
  publisher={APS}
}

@article{ozawa_topological_2019,
	title = {Topological photonics},
	volume = {91},
	doi = {10.1103/RevModPhys.91.015006},
	pages = {015006},
	number = {1},
	journal = {Rev. Mod. Phys.},
	author = {T. Ozawa and H.M. Price and A. Amo and N. Goldman and M. Hafezi and L. Lu and M.C. Rechtsman and D. Schuster and J. Simon and O. Zilberberg and I. Carusotto},
	year = {2019}
}

@article{lu_topological_2014,
	title = {Topological photonics},
	issn = {1749-4885},
	doi = {10.1038/nphoton.2014.248},
	pages = {821--829},
	volume = {8},
	journal = {Nature Photonics},
	author = {L. Lu and J.D. Joannopoulos and M. Soljačić},
	year = {2014}
}

@article{https://doi.org/10.1002/lpor.202100300,
author = {Tang, Guo-Jing and He, Xin-Tao and Shi, Fu-Long and Liu, Jian-Wei and Chen, Xiao-Dong and Dong, Jian-Wen},
title = {Topological Photonic Crystals: Physics, Designs, and Applications},
journal = {Laser \& Photonics Reviews},
volume = {16},
number = {4},
pages = {2100300},
doi = {https://doi.org/10.1002/lpor.202100300},
year = {2022}
}

@article{10.1063/5.0058478,
    author = {Schulz, Julian and Vaidya, Sachin and Jörg, Christina},
    title = "{Topological photonics in 3D micro-printed systems}",
    journal = {APL Photonics},
    volume = {6},
    number = {8},
    pages = {080901},
    year = {2021},
    issn = {2378-0967},
    doi = {10.1063/5.0058478}
}

@article{doi:10.1126/science.aar4005,
author = {Miguel A. Bandres  and Steffen Wittek  and Gal Harari  and Midya Parto  and Jinhan Ren  and Mordechai Segev  and Demetrios N. Christodoulides  and Mercedeh Khajavikhan },
title = {Topological insulator laser: Experiments},
journal = {Science},
volume = {359},
number = {6381},
pages = {eaar4005},
year = {2018},
doi = {10.1126/science.aar4005}
}

@article{doi:10.1126/science.aar4003,
author = {Gal Harari  and Miguel A. Bandres  and Yaakov Lumer  and Mikael C. Rechtsman  and Y. D. Chong  and Mercedeh Khajavikhan  and Demetrios N. Christodoulides  and Mordechai Segev },
title = {Topological insulator laser: Theory},
journal = {Science},
volume = {359},
number = {6381},
pages = {eaar4003},
year = {2018},
doi = {10.1126/science.aar4003}
}

@article{hafezi_robust_2011,
	title = {Robust optical delay lines with topological protection},
	volume = {7},
	issn = {1745-2473},
	doi = {10.1038/nphys2063},
	pages = {907--912},
	number = {11},
	journal = {Nat Phys},
	author = {Hafezi, Mohammad and Demler, Eugene A. and Lukin, Mikhail D. and Taylor, Jacob M.},
	year = {2011},
}

@article{vaidya2023reentrant,
  title={Reentrant delocalization transition in one-dimensional photonic quasicrystals},
  author={Vaidya, Sachin and J{\"o}rg, Christina and Linn, Kyle and Goh, Megan and Rechtsman, Mikael C},
  journal={Physical Review Research},
  volume={5},
  number={3},
  pages={033170},
  year={2023},
  publisher={APS}
}

@article{lustig2019edge,
  title={Photonic topological insulator in synthetic dimensions},
  author={Lustig, Eran and Weimann, Steffen and Plotnik, Yonatan and Lumer, Yaakov and Bandres, Miguel A and Szameit, Alexander and Segev, Mordechai},
  journal={Nature},
  volume={567},
  number={7748},
  pages={356--360},
  year={2019},
  publisher={Nature Publishing Group UK London}
}

@article{ozawa2019synthetic,
  title={Topological quantum matter in synthetic dimensions},
  author={Ozawa, Tomoki and Price, Hannah M},
  journal={Nature Reviews Physics},
  volume={1},
  number={5},
  pages={349--357},
  year={2019},
  publisher={Nature Publishing Group UK London}
}

@book{photoniccrystalsbook,
  author = {John D. Joannopoulos and Steven G. Johnson and Joshua N. Winn and Robert D. Meade},
  title = {Photonic Crystals: Molding the Flow of Light},
  publisher = {Princeton University Press},
  year = 2008,
  edition = {second},
  month = {February},
  url = {http://ab-initio.mit.edu/book}
}

@article{DUTRA2023114338,
title = {Hofstadter butterfly in optical multilayers},
journal = {Optical Materials},
volume = {144},
pages = {114338},
year = {2023},
issn = {0925-3467},
doi = {https://doi.org/10.1016/j.optmat.2023.114338},
author = {R.F. Dutra and M.S. Vasconcelos and D.H.A.L. Anselmo}
}

@article{PhysRevLett.132.266601,
  title = {Weyl Points on Nonorientable Manifolds},
  author = {Fonseca, Andr\'e Grossi and Vaidya, Sachin and Christensen, Thomas and Rechtsman, Mikael C. and Hughes, Taylor L. and Solja\ifmmode \check{c}\else \v{c}\fi{}i\ifmmode \acute{c}\else \'{c}\fi{}, Marin},
  journal = {Phys. Rev. Lett.},
  volume = {132},
  issue = {26},
  pages = {266601},
  numpages = {7},
  year = {2024},
  month = {Jun},
  publisher = {American Physical Society},
  doi = {10.1103/PhysRevLett.132.266601}
}

@article{peierls_zur_1933,
	title = {Zur {Theorie} des {Diamagnetismus} von {Leitungselektronen}},
	volume = {80},
	issn = {0044-3328},
	url = {https://doi.org/10.1007/BF01342591},
	doi = {10.1007/BF01342591},
	language = {de},
	number = {11},
	urldate = {2025-01-24},
	journal = {Zeitschrift für Physik},
	author = {Peierls, R.},
	month = nov,
	year = {1933},
	pages = {763--791},
}

@article{zhou2017optical,
  title={Optical isolation with nonlinear topological photonics},
  author={Zhou, Xin and Wang, You and Leykam, Daniel and Chong, Yi Dong},
  journal={New Journal of Physics},
  volume={19},
  number={9},
  pages={095002},
  year={2017},
  publisher={IOP Publishing}
}

@article{el2013chip,
  title={On-chip non-reciprocal optical devices based on quantum inspired photonic lattices},
  author={El-Ganainy, Ramy and Eisfeld, Alexander and Levy, Miguel and Christodoulides, Demetrios N},
  journal={Applied Physics Letters},
  volume={103},
  number={16},
  year={2013},
  publisher={AIP Publishing}
}

@article{el2015optical,
  title={Optical isolation in topological-edge-state photonic arrays},
  author={El-Ganainy, Ramy and Levy, Miguel},
  journal={Optics letters},
  volume={40},
  number={22},
  pages={5275--5278},
  year={2015},
  publisher={Optica Publishing Group}
}

@article{st2017lasing,
  title={Lasing in topological edge states of a one-dimensional lattice},
  author={St-Jean, Philippe and Goblot, Victor and Galopin, Elisabeth and Lema{\^\i}tre, A and Ozawa, Thomas and Le Gratiet, Luc and Sagnes, Isabelle and Bloch, Jacqueline and Amo, Alberto},
  journal={Nature Photonics},
  volume={11},
  number={10},
  pages={651--656},
  year={2017},
  publisher={Nature Publishing Group UK London}
}

@article{johnson2001block,
  title={Block-iterative frequency-domain methods for Maxwell’s equations in a planewave basis},
  author={Johnson, Steven G and Joannopoulos, John D},
  journal={Optics express},
  volume={8},
  number={3},
  pages={173--190},
  year={2001},
  publisher={Optical Society of America}
}

@article{aubry1980analyticity,
  title={Analyticity breaking and Anderson localization in incommensurate lattices},
  author={Aubry, Serge and Andr{\'e}, Gilles},
  journal={Ann. Israel Phys. Soc},
  volume={3},
  number={133},
  pages={18},
  year={1980}
}

@article{roy2022critical,
  title={Critical analysis of the reentrant localization transition in a one-dimensional dimerized quasiperiodic lattice},
  author={Roy, Shilpi and Chattopadhyay, Sourav and Mishra, Tapan and Basu, Saurabh},
  journal={Physical Review B},
  volume={105},
  number={21},
  pages={214203},
  year={2022},
  publisher={APS}
}

@article{padhan2022emergence,
  title={Emergence of multiple localization transitions in a one-dimensional quasiperiodic lattice},
  author={Padhan, Ashirbad and Giri, Mrinal Kanti and Mondal, Suman and Mishra, Tapan},
  journal={Physical Review B},
  volume={105},
  number={22},
  pages={L220201},
  year={2022},
  publisher={APS}
}

@article{landau1932theorie,
  title={Zur theorie der energieubertragung. II},
  author={Landau, Lev},
  journal={Physikalische Zeitschrift der Sowjetunion},
  volume={2},
  pages={46},
  year={1932}
}

@article{zener1932non,
  title={Non-adiabatic crossing of energy levels},
  author={Zener, Clarence},
  journal={Proceedings of the Royal Society of London. Series A, Containing Papers of a Mathematical and Physical Character},
  volume={137},
  number={833},
  pages={696--702},
  year={1932},
  publisher={The Royal Society London}
}

@article{xing2022quantum,
  title={Quantum transport in a one-dimensional quasicrystal with mobility edges},
  author={Xing, Yan and Qi, Lu and Zhao, Xuedong and L{\"u}, Zhe and Liu, Shutian and Zhang, Shou and Wang, Hong-Fu},
  journal={Physical Review A},
  volume={105},
  number={3},
  pages={032443},
  year={2022},
  publisher={APS}
}

@article{johnstone2022bulk,
  title={Bulk localized transport states in infinite and finite quasicrystals via magnetic aperiodicity},
  author={Johnstone, Dean and Colbrook, Matthew J and Nielsen, Anne EB and {\"O}hberg, Patrik and Duncan, Callum W},
  journal={Physical Review B},
  volume={106},
  number={4},
  pages={045149},
  year={2022},
  publisher={APS}
}

@article{hatsugai1993,
  title = {Chern number and edge states in the integer quantum Hall effect},
  author = {Hatsugai, Yasuhiro},
  journal = {Phys. Rev. Lett.},
  volume = {71},
  issue = {22},
  pages = {3697--3700},
  numpages = {0},
  year = {1993},
  month = {Nov},
  publisher = {American Physical Society},
  doi = {10.1103/PhysRevLett.71.3697},
  url = {https://link.aps.org/doi/10.1103/PhysRevLett.71.3697}
}

@article{thouless1983quantization,
  title={Quantization of particle transport},
  author={Thouless, DJ},
  journal={Physical Review B},
  volume={27},
  number={10},
  pages={6083},
  year={1983},
  publisher={APS}
}

@misc{yeh1990optical,
  title={Optical waves in layered media},
  author={Yeh, Pochi and Hendry, Michael},
  year={1990},
  publisher={American Institute of Physics}
}

@article{anderson1958absence,
  title={Absence of diffusion in certain random lattices},
  author={Anderson, Philip W and others},
  journal={Physical review},
  volume={109},
  number={5},
  pages={1492--1505},
  year={1958}
}

@article{stutzer2018photonic,
  title={Photonic topological Anderson insulators},
  author={St{\"u}tzer, Simon and Plotnik, Yonatan and Lumer, Yaakov and Titum, Paraj and Lindner, Netanel H and Segev, Mordechai and Rechtsman, Mikael C and Szameit, Alexander},
  journal={Nature},
  volume={560},
  number={7719},
  pages={461--465},
  year={2018},
  publisher={Nature Publishing Group UK London}
}

@article{li2009topological,
  title={Topological anderson insulator},
  author={Li, Jian and Chu, Rui-Lin and Jain, Jainendra K and Shen, Shun-Qing},
  journal={Physical review letters},
  volume={102},
  number={13},
  pages={136806},
  year={2009},
  publisher={APS}
}

@article{kraus2012topological,
  title={Topological states and adiabatic pumping in quasicrystals},
  author={Kraus, Yaacov E and Lahini, Yoav and Ringel, Zohar and Verbin, Mor and Zilberberg, Oded},
  journal={Physical review letters},
  volume={109},
  number={10},
  pages={106402},
  year={2012},
  publisher={APS}
}

@article{groth2009theory,
  title={Theory of the topological Anderson insulator},
  author={Groth, CW and Wimmer, M and Akhmerov, AR and Tworzyd{\l}o, J and Beenakker, CWJ},
  journal={Physical review letters},
  volume={103},
  number={19},
  pages={196805},
  year={2009},
  publisher={APS}
}

@article{PhysRevB.83.075105,
  title = {Localization in one-dimensional lattices with non-nearest-neighbor hopping: Generalized Anderson and Aubry-Andr\'e models},
  author = {Biddle, J. and Priour, D. J. and Wang, B. and Das Sarma, S.},
  journal = {Phys. Rev. B},
  volume = {83},
  issue = {7},
  pages = {075105},
  numpages = {22},
  year = {2011},
  month = {Feb},
  publisher = {American Physical Society},
  doi = {10.1103/PhysRevB.83.075105},
  url = {https://link.aps.org/doi/10.1103/PhysRevB.83.075105}
}

@article{PhysRevLett.104.070601,
  title = {Predicted Mobility Edges in One-Dimensional Incommensurate Optical Lattices: An Exactly Solvable Model of Anderson Localization},
  author = {Biddle, J. and Das Sarma, S.},
  journal = {Phys. Rev. Lett.},
  volume = {104},
  issue = {7},
  pages = {070601},
  numpages = {4},
  year = {2010},
  month = {Feb},
  publisher = {American Physical Society},
  doi = {10.1103/PhysRevLett.104.070601},
  url = {https://link.aps.org/doi/10.1103/PhysRevLett.104.070601}
}

@article{li2017mobility,
  title={Mobility edges in one-dimensional bichromatic incommensurate potentials},
  author={Li, Xiao and Li, Xiaopeng and Das Sarma, S},
  journal={Physical Review B},
  volume={96},
  number={8},
  pages={085119},
  year={2017},
  publisher={American Physical Society}
}

@article{PhysRevLett.126.106803,
  title = {Reentrant Localization Transition in a Quasiperiodic Chain},
  author = {Roy, Shilpi and Mishra, Tapan and Tanatar, B. and Basu, Saurabh},
  journal = {Phys. Rev. Lett.},
  volume = {126},
  issue = {10},
  pages = {106803},
  numpages = {5},
  year = {2021},
  month = {Mar},
  publisher = {American Physical Society},
  doi = {10.1103/PhysRevLett.126.106803},
  url = {https://link.aps.org/doi/10.1103/PhysRevLett.126.106803}
}

@article{PhysRevLett.103.013901,
  title = {Observation of a Localization Transition in Quasiperiodic Photonic Lattices},
  author = {Lahini, Y. and Pugatch, R. and Pozzi, F. and Sorel, M. and Morandotti, R. and Davidson, N. and Silberberg, Y.},
  journal = {Phys. Rev. Lett.},
  volume = {103},
  issue = {1},
  pages = {013901},
  numpages = {4},
  year = {2009},
  month = {Jun},
  publisher = {American Physical Society},
  doi = {10.1103/PhysRevLett.103.013901},
  url = {https://link.aps.org/doi/10.1103/PhysRevLett.103.013901}
}

@article{roati2008anderson,
  title={Anderson localization of a non-interacting Bose--Einstein condensate},
  author={Roati, Giacomo and D’Errico, Chiara and Fallani, Leonardo and Fattori, Marco and Fort, Chiara and Zaccanti, Matteo and Modugno, Giovanni and Modugno, Michele and Inguscio, Massimo},
  journal={Nature},
  volume={453},
  number={7197},
  pages={895--898},
  year={2008},
  publisher={Nature Publishing Group UK London}
}

@article{PhysRevLett.120.160404,
  title = {Single-Particle Mobility Edge in a One-Dimensional Quasiperiodic Optical Lattice},
  author = {L\"uschen, Henrik P. and Scherg, Sebastian and Kohlert, Thomas and Schreiber, Michael and Bordia, Pranjal and Li, Xiao and Das Sarma, S. and Bloch, Immanuel},
  journal = {Phys. Rev. Lett.},
  volume = {120},
  issue = {16},
  pages = {160404},
  numpages = {6},
  year = {2018},
  month = {Apr},
  publisher = {American Physical Society},
  doi = {10.1103/PhysRevLett.120.160404},
  url = {https://link.aps.org/doi/10.1103/PhysRevLett.120.160404}
}

@article{PhysRevLett.122.170403,
  title = {Observation of Many-Body Localization in a One-Dimensional System with a Single-Particle Mobility Edge},
  author = {Kohlert, Thomas and Scherg, Sebastian and Li, Xiao and L\"uschen, Henrik P. and Das Sarma, Sankar and Bloch, Immanuel and Aidelsburger, Monika},
  journal = {Phys. Rev. Lett.},
  volume = {122},
  issue = {17},
  pages = {170403},
  numpages = {7},
  year = {2019},
  month = {May},
  publisher = {American Physical Society},
  doi = {10.1103/PhysRevLett.122.170403},
  url = {https://link.aps.org/doi/10.1103/PhysRevLett.122.170403}
}

@article{maczewsky2020synthesizing,
  title={Synthesizing multi-dimensional excitation dynamics and localization transition in one-dimensional lattices},
  author={Maczewsky, Lukas J and Wang, Kai and Dovgiy, Alexander A and Miroshnichenko, Andrey E and Moroz, Alexander and Ehrhardt, Max and Heinrich, Matthias and Christodoulides, Demetrios N and Szameit, Alexander and Sukhorukov, Andrey A},
  journal={Nature Photonics},
  volume={14},
  number={2},
  pages={76--81},
  year={2020},
  publisher={Nature Publishing Group UK London}
}

@article{Mittal2014TopologicallyRobust,
  author       = {Mittal, S. and Fan, J. and Faez, S. and Migdall, A. and Taylor, J. M. and Hafezi, M.},
  title        = {Topologically Robust Transport of Photons in a Synthetic Gauge Field},
  journal      = {Physical Review Letters},
  volume       = {113},
  number       = {087403},
  year         = {2014},
  publisher    = {American Physical Society},
  doi          = {10.1103/PhysRevLett.113.087403}
}

@article{chang2013experimental,
  title={Experimental observation of the quantum anomalous Hall effect in a magnetic topological insulator},
  author={Chang, Cui-Zu and Zhang, Jinsong and Feng, Xiao and Shen, Jie and Zhang, Zuocheng and Guo, Minghua and Li, Kang and Ou, Yunbo and Wei, Pang and Wang, Li-Li and others},
  journal={Science},
  volume={340},
  number={6129},
  pages={167--170},
  year={2013},
  publisher={American Association for the Advancement of Science}
}

@article{yao2024observation,
  title={Observation of chiral edge transport in a rapidly rotating quantum gas},
  author={Yao, Ruixiao and Chi, Sungjae and Mukherjee, Biswaroop and Shaffer, Airlia and Zwierlein, Martin and Fletcher, Richard J},
  journal={Nature Physics},
  volume={20},
  number={11},
  pages={1726--1731},
  year={2024},
  publisher={Nature Publishing Group UK London}
}

@article{halperin1982quantized,
  title={Quantized Hall conductance, current-carrying edge states, and the existence of extended states in a two-dimensional disordered potential},
  author={Halperin, Bertrand I},
  journal={Physical review B},
  volume={25},
  number={4},
  pages={2185},
  year={1982},
  publisher={APS}
}

@article{klitzing1980new,
  title={New method for high-accuracy determination of the fine-structure constant based on quantized Hall resistance},
  author={Klitzing, K v and Dorda, Gerhard and Pepper, Michael},
  journal={Physical review letters},
  volume={45},
  number={6},
  pages={494},
  year={1980},
  publisher={APS}
}

@article{xing2021adiabatic,
  title={Adiabatic Pumping in a Generalized Aubry--Andr{\'e} Model Family with Mobility Edges},
  author={Xing, Yan and Qi, Lu and Zhao, Xuedong and L{\"u}, Zhe and Liu, Shutian and Zhang, Shou and Wang, Hong-Fu},
  journal={Annalen der Physik},
  volume={533},
  number={11},
  pages={2100270},
  year={2021},
  publisher={Wiley Online Library}
}
\clearpage

\setcounter{figure}{0}
\setcounter{table}{0}
\setcounter{equation}{0}
\renewcommand{\thefigure}{S\arabic{figure}}
\renewcommand{\thetable}{S\arabic{table}}
\renewcommand{\theequation}{S\arabic{equation}}

\onecolumngrid
\begin{center}
{\large\bfseries Supplemental Material for: Topological Pumping Through a Localized Bulk in a Photonic Hofstadter System\par}
\end{center}

\vspace{1.5em}
\twocolumngrid
\section{Methods}

\textbf{Simulation methods.} To compute the photonic eigenmodes, we use MPB in a one-dimensional supercell configuration. For each value of the swept parameter (\(\beta\) for butterfly or \(\phi\) for edge states), a multilayer stack is constructed explicitly from a sequence of unit cells, where each unit cell consists of one Si and one SiO\(_2\) layer with equal thickness. The complete stack of the unit cells (with additional cladding layers for edge states) forms the MPB supercell. The total physical thickness of this supercell defines the lattice constant used internally by MPB. MPB's spatial resolution parameter specifies the number of grid points per unit length in the coordinate system used to define the geometry. In other words, the total number of grid points is then just the resolution multiplied by the total length of the sample. We find that a resolution of 32 grid points per unit length, corresponding to approximately 32 grid points per unit cell on average, yields convergence in the IPR.  The refractive indices used for the constituent materials are \(n_{\text{Si}} = 3.54\) and \(n_{\text{SiO}_2} = 1.5\). 

For all spectra shown, the chosen initial thickness was \(t_0 = 0.5\). For all butterflies shown, since \(\phi\) is irrelevant, it is chosen to be \(\phi = 0\). For the low modulation cases, the amplitude was chosen to be \(A= 0.2\). For the high modulation cases, the amplitude was chosen to be \(A = 0.5\). The cladding layers used to image the chiral edge states each have an individual thickness of \(t_\text{clad} = 0.325\), and we choose an odd number of layers to pad each side so that the first layer in the stack is Si cladding and the last layer in the stack is SiO\(_2\) cladding. This preserves the alternating dielectric structure while preventing native oxide layers on the surface to reduce impedance mismatch to air in the experiment. For the transfer-matrix simulations, 1D PhCs are built similarly with the same indices of refraction, thicknesses, and cladding layers. For the plots of all the chiral edge states shown, we used 11 layers of cladding on each side of the 1D PhC, except for the results in Fig.~\ref{M-fig:fig 2}(b), (c) and Fig.~\ref{M-fig:fig 3}(f) which only have 3 layers of cladding since they include experimental results which require higher transmission signal.

\textbf{Experimental methods.} To fabricate these 1D PhCs in experiment, we scale the dimensionless thicknesses by an average lattice constant \(\braket{a}\). This quantity can be arbitrarily chosen as it only affects the wavelength range of the characterization method and fabrication times. We chose \(\braket{a} = 387.5\) nm for all results, except for Fig.~\ref{M-fig:fig 2}(b) and (c), which were smaller crystals at \(\braket{a} = 250\)~nm. The smaller crystals were fabricated first before we encountered issues with low signal, likely due to Si absorption at shorter wavelengths, and thus the rest of the samples were scaled up to avoid this lossy region. We deposit Si and SiO\(_2\) thin films onto a borosilicate Corning microscope cover glass with PECVD using the Applied Materials P5000 Cluster Tool. The tool reacts precursor gases under a controlled temperature, pressure, and applied RF power. The precursor gases used to deposit Si were SiH\(_4\) (90 sccm) and Ar (2000 sccm) at \SI{220}{\degree{C}}, 4.5 Torr and 100 W. SiO\(_2\) was deposited by mixing SiH\(_4\) (20 sccm), N\(_2\)O (840 sccm), and N\(_2\) (1400 sccm) at \SI{220}{\degree{C}}, 3.5 Torr, and 300 W. 

The resulting 1D PhCs were then characterized using an Agilent Cary 5000 UV-Vis-NIR spectrophotometer, which can measure transmission through a substrate over a wide wavelength range from 200 nm to 2500 nm by utilizing a monochromator illuminated by two lamps: a tungsten-halogen for visible and near-IR and a deuterium arc lamp for UV. All results were measured at normal incidence from 900 nm to 1900 nm, except for Fig.~\ref{M-fig:fig 2}(c) which was measured from 600 to 1200 nm. A wavelength step size of 1 nm and a spectral bandwidth of 4 nm were used. For each measured sample, the transmission spectra were normalized to a bare Corning microscope cover glass.

\section{High system size transmission of localization}

\begin{figure}[h]
    \centering
    \includegraphics[width=1\linewidth]{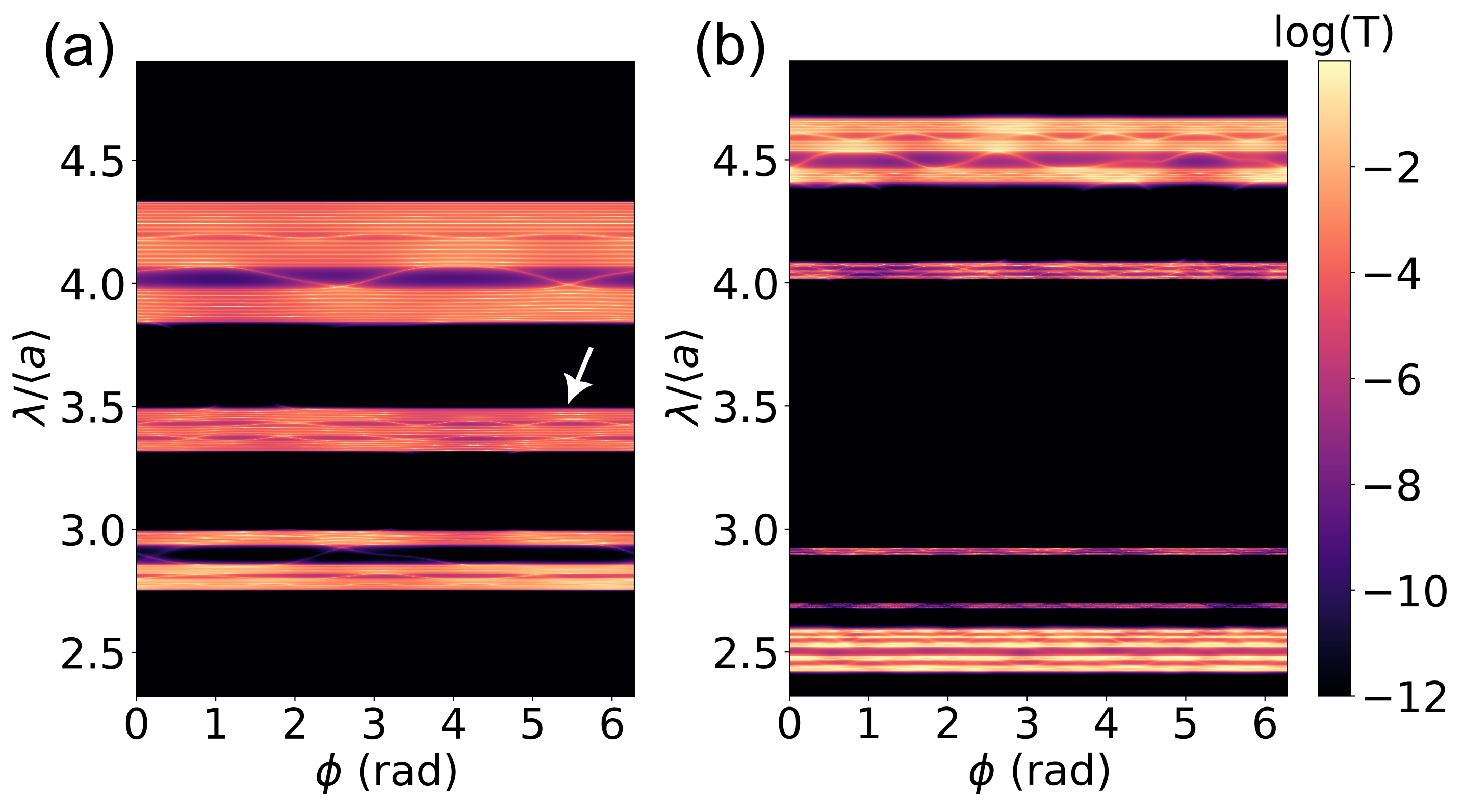}
    \caption{\small{Transfer-matrix simulation of logarithmic transmission at \(N=89\) and \(\beta = 144/89\) of the chiral edge states at (a) low modulation \(A=0.2\) and (b) high modulation \(A=0.5\). The bulk bands indicated by the white arrow disappear upon high quasiperiodic modulation.}}
    \label{supfig:1}
\end{figure}

To confirm whether the Chern bands from Fig.~\ref{M-fig:fig 3}(c) are indeed fully localized, we calculate through transfer-matrix simulations the transmission of chiral edge states at high system size. We choose a system size of \(N=89\) and probe for the logarithmic (base 10) transmission to detect possible faint signals. In Fig.~\ref{supfig:1}(a), which demonstrates the low modulation case at \(A=0.2\), all expected bulk bands from MPB are present. Because of their short localization lengths, the chiral edge states themselves are noticeably absent at this system size. The bulk bands indicated by the white arrow, near \(\lambda/\braket{a} \sim 3.5\) are expected to localize upon increasing quasiperioidc modulation. In Fig.~\ref{supfig:1}(b), the high modulation case at \(A =0.5\), we see those bulk bands vanish. Meanwhile, other extended bands are still present and are pushed apart in wavelength from the modulation. More mini-gaps in the surrounding bulk bands open as the system becomes more quasiperiodic, as a result of the fractal nature of the Hofstadter butterfly.

\begin{figure*}[]
    \centering
    \includegraphics[width=\textwidth]{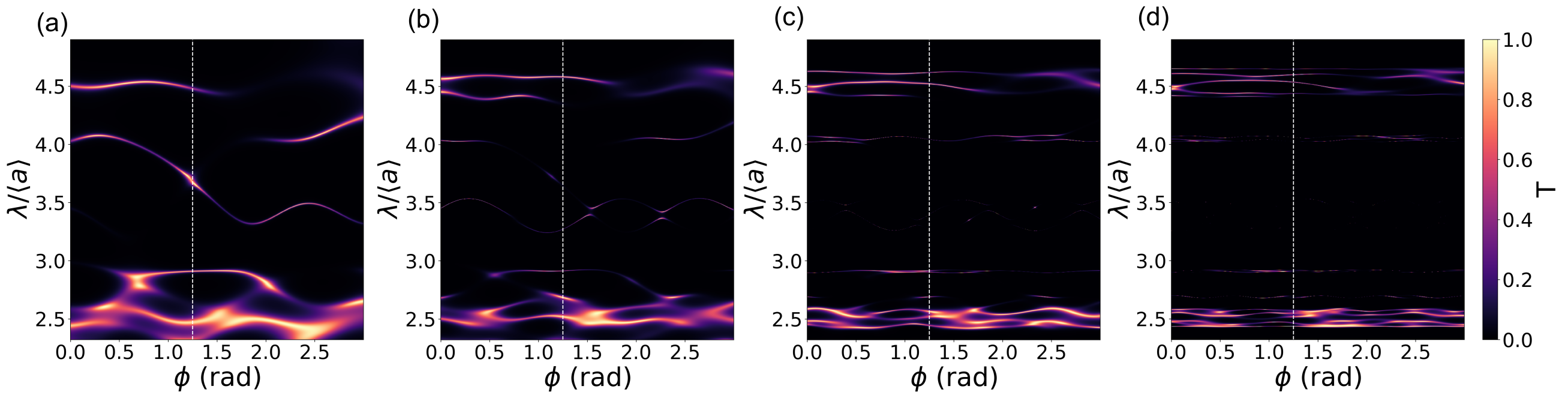}
    \caption{\small{The transfer-matrix simulations at: (a) \(N=5\) and \(\beta = 8/5\) (b) \(N=8\) and \(\beta = 13/8\) (c) \(N=13\) and \(\beta =21/13\) (d) \(N=21\) and \(\beta = 34/21 \), which shows progressively less transmissive bulk bands and edge states as the system size increases, implying localization. The white dashed line indicates the slice \(\phi = 1.25\) which was used in the experiment in the main text to diagnose the localization transition.}}
    \label{supfig:2}
\end{figure*}

\section{Edge state spectra of experimental slices}

The experimental data in the main text used in Fig.~\ref{M-fig:fig 3}(f) is the transmission spectrum at a single \(\phi  =1.25\) for various PhCs at different system sizes. We show the simulated spectra for these same PhCs at a larger range of \(\phi\) to corroborate them with the experiment. These simulated spectra are shown in Fig.~\ref{supfig:2}, which show the chiral edge states and bulk bands as they localize for system sizes of \(N=5, 8, 13, 21\). The white dashed line indicates \(\phi=1.25\), the slice used for the experimental data. Fig.~\ref{supfig:2}(a) shows the lowest system size at \(N=5\). Here, only the chiral edge state can be seen, which starts at \(\lambda/\braket{a} \sim 4.0\) and winds down. Due to the small system size, the spectrum has not yet converged to the bulk limit, with only the edge branch clearly visible and bulk bands still missing. As the system size increases to \(N=8\) (in Fig.~\ref{supfig:2}(b)), a faint bulk band appears below the chiral edge state, but both this band and the edge state are discernibly dimmer compared to \(N=5\). This indicates the onset of localization, especially since the bulk bands at the outer edges of the wavelength range, predicted to be extended by MPB, remain highly transmissive. This trend continues with the higher system sizes of \(N=13\) and \(N=21\) (Fig.~\ref{supfig:2}(c) and Fig.~\ref{supfig:2}(d) respectively), as middle bulk bands and edge states continue to drop in transmission where localization has fully set in.

\section{Video sweep of modulation strength in butterfly and edge states}

In the attached ancillary videos, we show the localization behavior of the photonic Hofstadter butterfly and its chiral edge states as the strength of the quasiperiodic modulation \(A\) is increased. The video depicting the Hofstadter butterfly is calculated for a system size of \(N = 89\), and as mentioned in the main text, has several regions which begin to localize around \(A \sim 0.5\). 

Strictly speaking, the Hofstadter butterfly is constructed from rational values of \(\beta\), for which the system is periodic (rather than quasiperiodic) and therefore does not undergo a true Aubry-Andr\'e localization transition. However, for a finite system, high-order rational approximants provide an accurate representation of the corresponding irrational modulation over the available length scale. As a result, the states that appear localized in these plots should be understood as finite-size manifestations of the quasiperiodic localization physics. Thus, although the \(\beta\) values are rational, they approximate the irrational case closely enough that the associated eigenstates display localization behavior over the finite sample.

As the quasiperiodic modulation increases from zero to \(A \sim 0.5\), extended bulk bands are pushed apart and mini-gaps open within the spectrum. Upon further increasing the modulation, these bands start to close in until at \(A \sim 0.8\), the aforementioned localized bands will begin to delocalize. This is known as a reentrant delocalization transition, which has been observed previously in this specific photonic crystal platform \cite{vaidya2023reentrant}.

A similar story occurs with the video of the chiral edge states, calculated for \(N=55\), \(\beta = 89/55\), and with 11 layers of cladding per side. Here we note the persistence of the edge states despite the localization of the Chern bands (near \(\lambda/\braket{a} \sim 3.5\)) at \(A\sim0.5\), as the gap remains open through the entire process. The same aforementioned reentrant transition also occurs at \(A \sim 0.8\).

The quoted values of \(A\sim 0.5\) should not be interpreted as a unique critical point for the full spectrum, since different spectral regions localize over different ranges of \(A\), consistent with mobility edge behavior. Rather, \(A\sim 0.5\) is the value at which the particular Chern band highlighted here is already fully localized. Owing to finite-size effects, the precise onset of localization is not sharply resolved in the present calculations; \(A\sim 0.5\) was therefore chosen as a representative value sufficiently beyond the localization threshold for the band of interest. The same could be said for the reentrant transition, and the exact critical values of \(A\) associated with these transitions could in principle be determined numerically through finite-size scaling.

\section{Tight-binding Review and Inspiration}

\begin{figure}[]
    \centering
    \includegraphics[width=\linewidth]{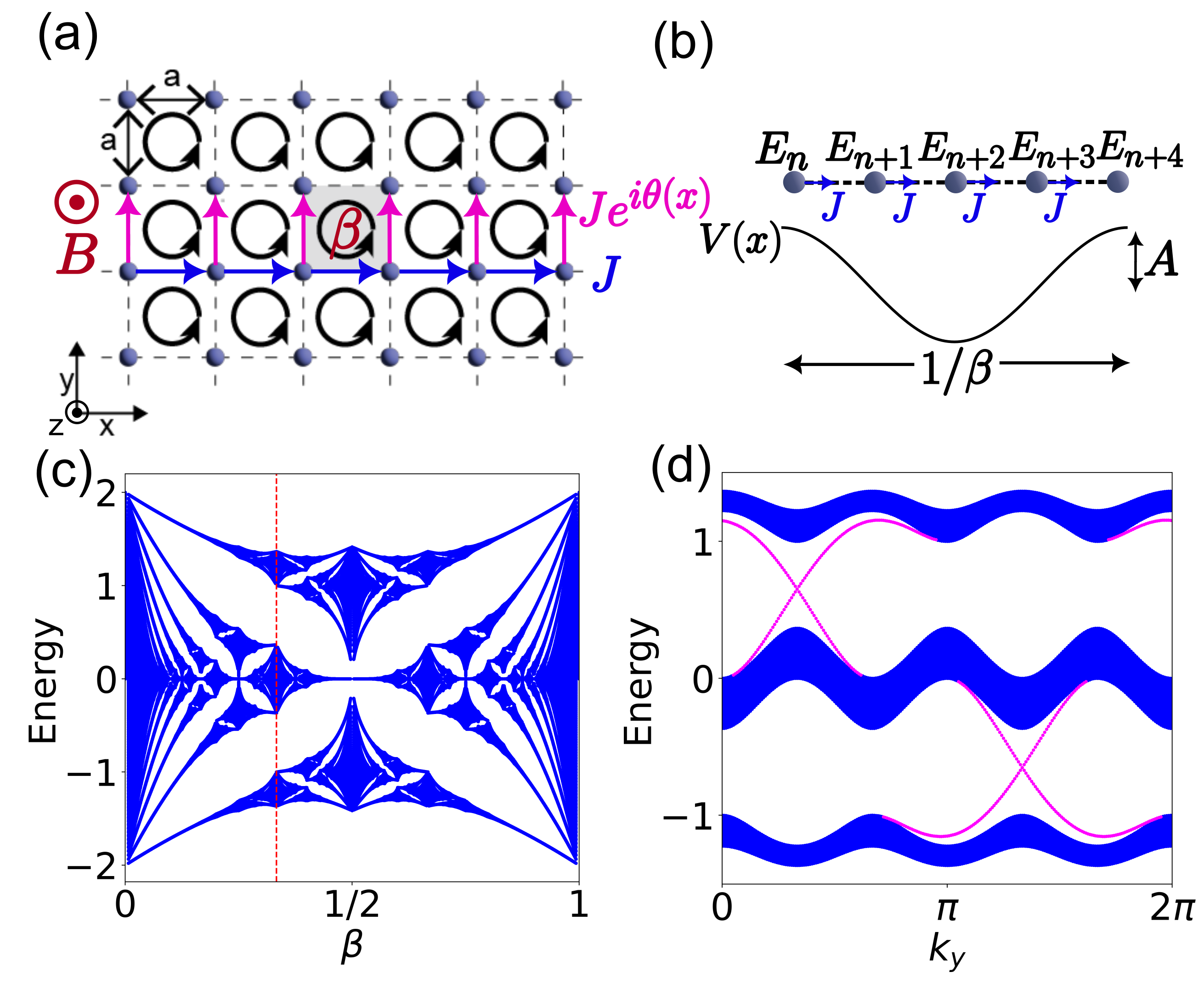}
    \caption{ \small{(a) 2D electron square lattice model with nearest neighbor couplings \(J\) and uniform magnetic field \(\mathbf{B} = B\hat{z}\). A magnetic flux \(\beta\) threads each plaquette, modifying the hoppings along \(y\) with an additional phase \(e^{i\theta(x)}\) breaking translational symmetry in the \(x\) direction. (b) The 2D lattice is reduced to a 1D tight-binding chain with potential \(V(x)\), couplings \(J\), and on-site energies $E_n=E_0 +A\cos(2\pi \beta n + k_y)$  with frequency \(\beta\) and amplitude \(A\). (c) The Harper-Hofstadter fractal butterfly obtained by diagonalizing Eq.~\eqref{harper eq} and sweeping for rational \(\beta\) for \(E_0=0\), \(A=1\), and \(J=0.5\). A red dashed line at \(\beta = 1/3\) indicates the slice of the butterfly chosen to reveal chiral edge states. (d) Chiral edge states for \(\beta = 1/3\) appear in the bandgaps between the three sliced bulk bands in (c), obtained by diagonalizing Eq.~\eqref{harper eq} and sweeping for \(k_y\) with open boundary conditions with the same parameters as (c).
      \label{supfig:3}}}
\end{figure}

The Hofstadter model consists of a two-dimensional square lattice with lattice constant \(a\) along the \((xy)\) plane, where lattice sites are coupled to their nearest neighbors, shown in Fig.~\ref{supfig:3}(a). Here, a magnetic field is applied uniformly normal to the lattice. The electron experiences a phase shift as it hops around a plaquette (the area enclosed by lattice points), exactly like the Aharonov-Bohm effect. In the Landau gauge, the hoppings along the \(y\)-direction are modified by an additional phase, known as a Peierls phase \cite{peierls_zur_1933}. This phase relates to the magnetic flux per plaquette and has a dependence on the \(x\)-direction. As a result, periodicity in the \(x\)-direction is broken, while discrete translational invariance remains in the \(y\)-direction. Bloch solutions in this \(y\)-direction reduce the 2D Hamiltonian into a 1D equation, known as the Harper (or Aubry-Andr\'e-Harper) equation \cite{hofstadter1976energy}. The problem can thus be reduced to a 1D tight-binding chain, shown in Fig.~\ref{supfig:3}(b), and the resulting Hamiltonian can be written as:

\begin{equation} \label{harper eq}
\begin{split}
       \hat{H} = & \sum_n \left[E_0+A\cos(2\pi \beta n + k_y)\right]\hat{c}^\dagger_n \hat{c}_n \\
    &-J \sum_n \left(\hat{c}^\dagger_{n+1}\hat{c}_n +\hat{c}^\dagger_{n} \hat{c}_{n+1}\right)  
\end{split}
\end{equation}
where \(\hat{c}^\dagger_n\) and \(\hat{c}_n\) are the creation and annihilation operators at lattice site \(n\) along the \(x\) direction, \(\beta\) is the magnetic flux per plaquette, \(E_0\) the on-site energy scale, \(J\) the hopping constant along the \(x\) direction, and \(k_y\) is the crystal momentum in the \(y\) direction. In the main text, we pattern each PhC using Eq.~\eqref{M-thickness eq}, which follows a similar form of the on-site energies in Eq.~\eqref{harper eq}. Comparing the two equations, there is then a mapping between \(\phi\), the synthetic dimension, and the transverse momentum \(k_y\) in the tight-binding case, both manifesting as a phase shift to the on-site modulation. \(\beta\) meanwhile in the photonic case represents a spatial frequency of modulation, playing the same role as the magnetic flux per plaquette in the tight-binding case as well.

To restore periodicity (and Bloch's theorem) in the \(x\)-direction, a supercell needs to form, known as a magnetic unit cell. If we write the magnetic flux per plaquette as a ratio of integers \(\beta=p/q\), then the lattice becomes periodic once again if we tile the lattice with a magnetic unit cell consisting of \(q\) subcells and thus \(q\) bands of eigenstates. However, \(\beta\) is a continuous parameter and can be irrational. If so, such cases are instead understood through ever-larger rational approximants to \(\beta\) with increasing \(q\). This causes a splitting in the bands which results in a fractal pattern known as the Hofstadter butterfly, where the energy eigenvalues as a function of magnetic flux \(\beta\) create a butterfly-like spectrum. Figure \ref{supfig:3}(c) depicts this fractal pattern, solved by diagonalizing a Hamiltonian with on-site energies set by Eq.~\eqref{harper eq} and choosing a fixed on-site modulation amplitude of \(A\) to cleanly resolve the butterfly's gaps.

It was also discovered that within the gaps of the butterfly, there exists a constant Hall conductance proportional to the Chern number \cite{thouless1982quantized}. By bulk-edge correspondence, each gap in the Hofstadter butterfly has an associated Chern number and topologically protected chiral edge states that live within it \cite{hatsugai1993}. The edge states can be seen by choosing a slice of \(\beta\) in the butterfly and calculating the band structure for open boundary conditions along the \(x\) direction. This is done by plotting the system's energy eigenvalues as a function of \(k_y\), shown in Fig. \ref{supfig:3}(d). Chiral edge states appear here as dispersive eigenstates (colored in pink) traversing the bandgap, with the two branches in each gap corresponding to the states on opposite physical edges. These edge states physically live in a 2D system, being confined to the finite edges of the \(x\) direction and flowing along the periodic \(y\) direction.

\end{document}